\documentclass[12pt]{article}
\usepackage{amsmath,amssymb,amscd}
\usepackage{epsfig}
\begin{document}


\begin{flushright}
\end{flushright}

\bigskip
\bigskip

\begin{center}
{\bf MARKET MILL DEPENDENCE PATTERN IN THE STOCK MARKET: DISTRIBUTION GEOMETRY, MOMENTS AND GAUSSIZATION}
\end{center}

\bigskip

\begin{center}
\bf{ \large Andrei Leonidov$^{(a,b,c)}$\footnote{Corresponding author. E-mail leonidov@lpi.ru}, Vladimir Trainin$^{(b)}$, \\Alexander
Zaitsev$^{(b)}$, Sergey Zaitsev$^{(b)}$}
\end{center}
\medskip
(a) {\it Theoretical Physics Department, P.N.~Lebedev Physics Institute,\\
    Moscow, Russia}

(b) {\it Letra Group, LLC, Boston, Massachusetts, USA}

(c) {\it Institute of Theoretical and Experimental Physics, Moscow, Russia}

\bigskip

\bigskip

\begin{center}
{\bf Abstract}
\end{center}

This paper continues a series of studies devoted to analysis of the bivariate probability distribution ${\cal P}(x,y)$ of two consecutive price
increments $x$ (push) and $y$ (response) at intraday timescales for a group of stocks. Besides the asymmetry properties of ${\cal P}(x,y)$ such as
Market Mill dependence patterns described in preceding paper \cite{LTZZ05a}, there are quite a few other interesting geometrical properties of this
distribution discussed in the present paper, e.g. transformation of the shape of equiprobability lines upon growing distance from the origin of $xy$
plane and approximate invariance of ${\cal P}(x,y)$ with respect to rotations at the multiples of $\pi/2$ around the origin of $xy$ plane. The
conditional probability distribution of response ${\cal P}(y|x)$ is found to be markedly non-gaussian at small magnitude of pushes and tending to
more gauss-like behavior upon growing push magnitude. The volatility of ${\cal P} (y|\,x)$ measured by the absolute value of the response shows
linear dependence on the absolute value of the push, and the skewness of ${\cal P} (y|\,x)$ is shown to inherit a sign of the push. The conditional
dynamics approach applied in this study is compared to regression models of AR-ARCH class.

\newpage

\section{Introduction}

Intensive investigations over recent decades have revealed statistically significant deviations from an assumption of independent identically
distributed (IID) increments that underlies the random walk model of stock price dynamics. A direct statistical evidence showing significant
deviations from IID using a BDS test \cite{BDS87} was presented in \cite{H91}. The rejection of the IID hypothesis also follows from the volatility -
based test described in \cite{Lo}. There is a number of dependence patterns showing themselves in such stylized facts as statistically significant
autocorrelations at intraday timescales, volatility clustering, leverage effect, etc. \cite{Lo,Man,SH,BP}, as well as correlations between
simultaneous increments of different stocks. Each of these effects corresponds to some sort of probabilistic dependence between the lagged and/or
simultaneous price increments and their moments.

In a series of papers \cite{LTZZ05a,LTZ05,LTZZ05b} including the present one we study further evidence of the presence of dependence patterns in
financial time series. An approach we apply is based on the direct analysis of the multivariate probability distributions of simultaneous and lagged
price increments for both single stocks and a basket of stocks. A simplest case we concentrate upon is that of the bivariate distribution describing
the interdependence of two price increments in two coinciding or non-overlapping time intervals. We also consider a natural and transparent
interpretation of the bivariate probability distribution provided by its sections corresponding to the fixed value of one of the variables, i.e.
conditional distributions.


Despite their fundamental importance, multidimensional probability distributions of stock price returns (increments) are, to the best of our
knowledge, not widely used. The bivariate distribution of returns in two consecutive intervals was analyzed, in the particular case of Levy-type
marginals, in \cite{M63}, where some interesting geometric features of this distribution both for the case of independent and dependent returns were
described. As discussed in \cite{MB}, the bivariate distribution in question can be considered as a "fingerprint" reflecting the nature of the
pattern embracing the two consecutive returns. Let us also mention the "compass rose" phenomenon \cite{CL96,AV04,V04} and the discussion of return
predictability in \cite{CL96} and \cite{V04}. Another line of research is an explicit analytical reconstruction of the bivariate distribution in
question using copulas, see e.g. \cite{EMS99,ELM01,MS01,JR01}. There is a few studies devoted to a direct analysis of the conditional distributions.
Recently an analysis of volatility dynamics exploiting such conditional distributions was described in \cite{CJY05}. The first moment of the
corresponding conditional distribution for daily time intervals was studied in \cite{BM03}.

At the same time the main focus of the efforts to quantify the conditional dynamics of financial instruments was on constructing and studying the
regression models, see e.g. \cite{SH,TL80,E82,EB86,B86,B87,H94,HS99,JR00,LL02,CD04}. Each of these models realizes a particular version of the
conditional dynamics, in which parameters of the conditional distribution describing the forthcoming increment depend on lagged increments and
moments.  In the simplest version of ARCH model \cite{SH,E82} this conditional distribution is gaussian with a standard deviation depending on the
magnitude(s) of one or more lagged increments. In GARCH models \cite{SH,EB86} the conditional standard deviation depends on lagged standard deviation
as well. As such conditionally gaussian approach did not allow to describe an observed degree of fat-tailedness of the increments, further
development included considering fat-tailed conditional distributions \cite{B87} and, in modern versions \cite{HS99,JR00,LL02,CD04}, the fat-tailed
and skewed conditional distributions in which fat-tailedness and skewness are conditional as well. A case of nonlinear dependence of the conditional
mean of forthcoming increment on lagged increments could most naturally be treated within a class of threshold autoregressive models \cite{TL80}.

An approach based on constructing regression type models is, by definition, a parametric one. Assuming some specific form of increment dynamics one
runs statistical tests to determine optimal values of the parameters in question. Our approach \cite{LTZZ05a,LTZ05,LTZZ05b} is, on contrary,
inherently non-parametric. We do not use any assumptions on the particular form of probabilistic links between price increments. The analysis of
price dynamics is made in terms of direct examination of the observed multivariate probability distributions.

In our study of dependence patterns in stock price dynamics \cite{LTZ05} a direct examination of the moments of the bivariate distributions linking
consecutive returns of a stock and simultaneous returns of a pair of stocks was performed. It was shown that some empirical features of the bivariate
distribution in question, e.g. conditional volatility smile, result from non-gaussian nature of the distirbution.

In the preceding paper \cite{LTZZ05a} we analyze the asymmetry properties of the bivariate probability distribution ${\cal P}(x,y)$ of two
consecutive stock price increments $x$ (push) and $y$ (response) resulting in remarkable market mill dependence patterns. In the present paper we
continue the analysis of \cite{LTZZ05a,LTZ05} by a more close inspection of the properties of the full bivariate distribution ${\cal P}(x,y)$ of
increments and the conditional distribution of the response ${\cal P}(y |\, x)$. Therefore, despite the fact that we do not discuss specifically the
market mill properties in the present paper, this paper still belongs to the series of studies \cite{LTZZ05a,LTZ05,LTZZ05b} under the umbrella title
- "Market mill dependence pattern ...".

The paper is organized as follows. In paragraph 2.1 we describe the dataset and the probabilistic methodology used in our analysis.

A detailed description of the results is given in paragraph 2.2. In 2.2.1 we study geometrical properties of the full bivariate probability
distribution ${\cal P}(x,y)$. First we show that the shape of the equiprobability lines of the distribution ${\cal P}(x,y)$ changes upon growing
distance from the origin of $xy$ plane. Another property of ${\cal P}(x,y)$ is its approximate invariance with respect to rotations at the multiple
of $\pi/2$ around the origin of $xy$ plane. In 2.2.2 we switch to studying properties of the conditional probability distribution ${\cal P}(y|\,x)$
and find that the volatility of ${\cal P}(y|\,x)$ measured by the absolute value of the response shows linear dependence on the absolute value of the
push, and the skewness of ${\cal P}(y|\,x)$ inherits a sign of the push.

In the discussion section of the paper we first concentrate on non-gaussian properties of both the full bivariate distribution ${\cal P}(x,y)$ and
conditional one ${\cal P}(y|\,x)$ (see paragraph 3.1). We discuss that the conditional distribution ${\cal P}(y|\,x)$ tends to more gauss-like
behavior upon growing magnitude of the push.

Our studies of the bivariate distribution are, by default, also studies of a special version of the conditional dynamics in which information on the
value of a price increment is fully accounted for in describing the probabilistic pattern characterizing the next increment. It is therefore of
direct interest to compare our results with the ones obtained within the regression model approach. This is done in section 3.2 of the paper.

The paragraph 4 finalizes the paper by describing main conclusions and outlook of the future studies.

\section{Properties of push - return distribution}

An analysis of the properties of the push - return distribution described in the present paragraph continues the study initiated in \cite{LTZZ05a}.
It goes into further details in describing the unique properties of the geometry of the bivariate probability distribution under consideration and
analyzes moments of the corresponding conditional distribution \footnote{Although an analysis of the properties of the bivariate probability
distribution and the corresponding conditional distribution presented in the present paper is self-contained, we strongly refer to \cite{LTZZ05a} for
a more comprehensive picture.}.

\subsection{Data and methodology}

Our study of high frequency dynamics of stock prices is based on data on the prices of 100 stocks traded in NYSE and NASDAQ in 2003-2004 sampled at 1
minute frequency\footnote{The list of stocks is given in the Appendix}.

Let us consider two non-overlapping time intervals of length $\Delta T_1$ and $\Delta T_2$, where the interval $\Delta T_2$ immediately follows after
$\Delta T_1$. We shall denote the price increment in the first interval $p(t_1 + \Delta T_1)-p(t_1)$ (push) by $x$ and that in the second one $p(t_2
+ \Delta T_2)-p(t_2)$ (response) by $y$. In this study we consider $\Delta T_1 = \Delta T_2 = $ 1 min., 3 min., 6 min.  The full probabilistic
description in $xy$ plane is given by the bivariate probability density ${\cal P} (x,y)$. An analysis of the full bivariate distribution ${\cal
P}(x,y)$ is often facilitated by considering its cross-sections corresponding to conditional distributions such as, e.g., the conditional
distribution of response at given push ${\cal P}(y|\,x)={\cal P}(x,y)/{\cal P}(x)$.

Let us stress that our data set combines pairs of price increments for all stocks belonging to the specified group
(see Appendix), so that a set of events (pairs of increments) unifies all subsets of events characterizing
individual stocks. A further detailed study of general features of the probability distributions ${\cal P} (x,y)$
and ${\cal P}(y|\,x)$ constitutes the main topic of the present paper and comprehends the analysis of
\cite{LTZZ05a} .

Let us note that analogously to \cite{LTZZ05a} in our analysis we use price increments. The results obtained using returns are in qualitative
agreement to the ones discussed in the present paper.

\subsection{Structure of the bivariate distribution}

Let us discuss properties of the bivariate distribution ${\cal P} (x,y)$.

\subsubsection{Global two-dimensional geometry. Invariance with respect to rotations.}

The two-dimensional projection of $\log_8 {\cal P} (x,y)$ in case of two adjacent 1-minute intervals is shown
\begin{figure}[h]
 \begin{center}
 \leavevmode
 \epsfysize=11cm
 \epsfbox{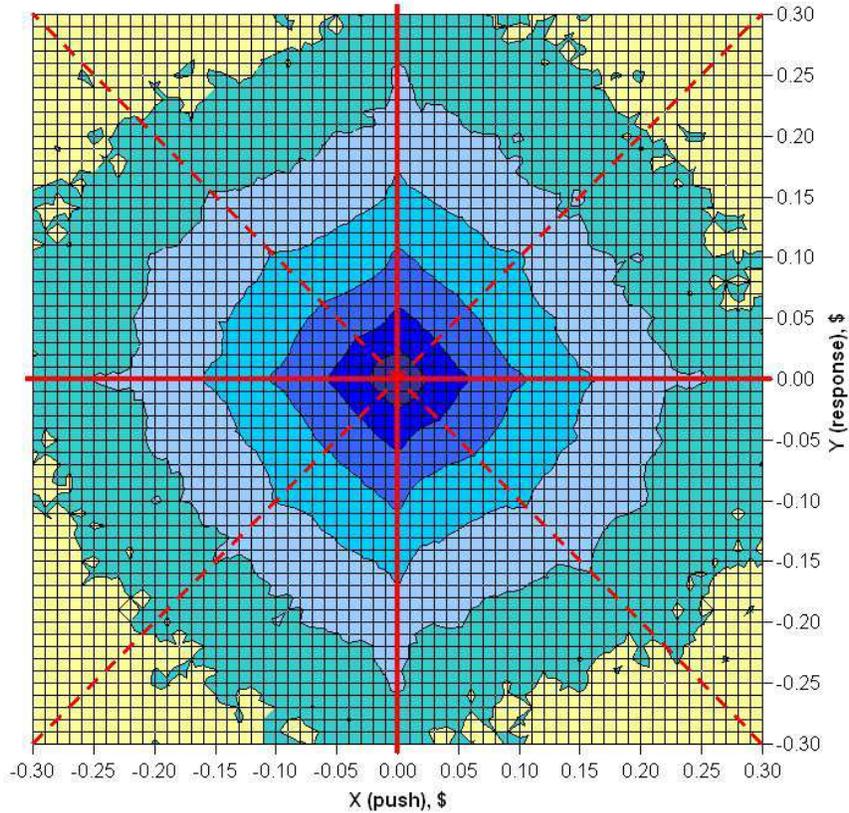}
 \end{center}
\caption{Logarithm of two-dimensional distribution $\log_8 {\cal P}(x,y)$, $\Delta T=1$ min. }
\end{figure}
in Fig.~1. To facilitate a discussion of some qualitative features seen in Fig.~1, let us sketch profiles of the equiprobability levels  of ${\cal P}
(x,y)$ in Fig.~2,
\begin{figure}[h]
 \begin{center}
 \leavevmode
 \epsfysize=7cm
 \epsfbox{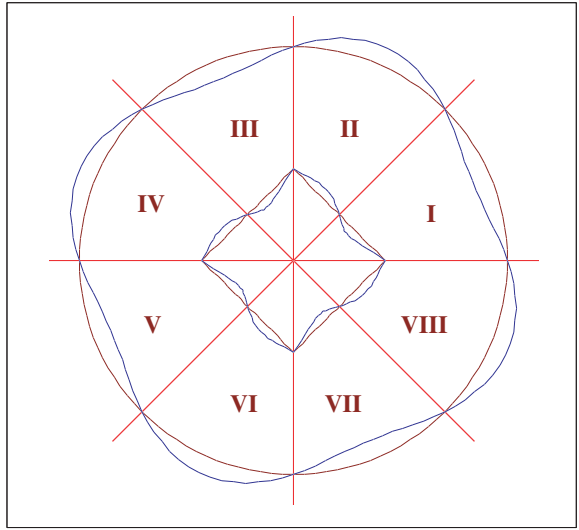}
 \end{center}
\caption{Sketch of the equiprobability levels of the bivariate distribution ${\cal P} (x,y)$. The basic regular symmetric structure is shown by brown
lines, the actual structure - by blue ones. }
\end{figure}
where the $xy$ plane is divided into sectors numbered counterclockwise from I to VIII. The shape of equiprobability lines shown in Figs.~1,2 can be
described as a superposition of a basic regular pattern, rhomboid in the vicinity of the origin and circular away from it, perturbed in such a way
that each of the even sectors (II,IV,VI,VIII) contains more probability density than each of the odd ones (sectors I,III,V,VII). The geometry of
unperturbed basic regular pattern (shown in brown in Fig.~2) can be described as
\begin{equation}\label{equil}
|x|^\alpha + |y|^\alpha \, = \, {\rm const} \, ,
\end{equation}
where $\alpha  \sim 1$ in the vicinity of the origin  and $\alpha \sim 2$ far away from it.

An interesting property of the bivariate distribution ${\cal P}(x,y)$ is its approximate invariance with respect to rotations at the multiples of
$\pi/2$. The distribution geometry is quite nontrivial: as already mentioned, all even sectors contain more probability density than the odd ones
\footnote{The nontrivial asymmetric properties of the distribution ${\cal P}(x,y)$ leading to the market mill structure and z-shaped structure of the
conditional mean response was analyzed in full details in \cite{LTZZ05a}. }. In terms of sample paths (pairs of increments) composed by increments
$\pm \zeta_1$ and $\pm \zeta_2$ the exact symmetry with respect to rotations at multiples of $\pi/2$ leads to the following chain of equalities:
\begin{equation}\label{rotations}
{\cal P}(\zeta_1,\zeta_2) = {\cal P}(-\zeta_2,\zeta_1) = {\cal P}(-\zeta_1, - \zeta_2) = {\cal P}(\zeta_2,-\zeta_1) \, .
\end{equation}
An approximate character of the symmetry with respect to rotations at multiples of $\pi/2$ shows itself in varying degree of proximity of the
corresponding distributions. To give a quantitative estimate of this proximity we consider three bivariate probability distributions ${\cal
P}^{\pi/2} (x,y)$, ${\cal P}^{\pi} (x,y)$ and ${\cal P}^{3 \pi/2} (x,y)$ obtained by rotating the original distribution ${\cal P}(x,y)$ at an angle
$\phi_i = i \cdot \pi/2$, where $i=1,2,3$. To compute a distance between two matrices corresponding to distributions ${\cal P}^{\phi_i}(x,y)$ and
${\cal P}^{\phi_j}(x,y)$ obtained by rotations of ${\cal P}(x,y)$ at $\phi_i$ and $\phi_j$ respectively we flatten each of them into a vector,
normalize it and compute a distance $D_1(\phi_i,\phi_j) \equiv {\rm dist}_{L_1} ({\cal P}^{\phi_i},{\cal P}^{\phi_j})$ between these vectors using
the $L_1$ ("Manhattan") metric\footnote{In this estimate we restrict our consideration to the domain $ \{ |x| \leq \$ \,\, 0.3, |y| \leq \$ \,\, 0.3
\}$ .}. We find
\begin{eqnarray}
 D_1(0,\pi/2) & = & D_1(0,3\pi/2) \, = \, D_1(\pi/2,3\pi/2) \, = \, D_1(\pi,3\pi/2) \\
 D_1(0,\pi) & = & D_1(\pi/2,3\pi/2)
\end{eqnarray}
and
\begin{equation}
 \frac{D_1(0,\pi)}{D_1(0,\pi/2} \, = \, \frac{D_1(\pi/2,3\pi/2)}{D_1(\pi/2,\pi)} \, = \, 0.34
\end{equation}
Therefore we see that the rotation at $\pi$ is a "better" symmetry of the full distribution than the rotation at $\pi/2$ implying, in turn, that
equality ${\cal P}(\zeta_1,\zeta_2) =  {\cal P}(-\zeta_1, - \zeta_2)$ holds to a better accuracy than ${\cal P}(\zeta_1,\zeta_2) =  {\cal
P}(-\zeta_2,\zeta_2)$.

\subsubsection{Geometry of response profile}

A detailed view on the distribution ${\cal P}(x,y)$  is provided by examining the corresponding conditional distributions such as, e.g., ${\cal
P}(y|x)={\cal P}(x,y)/{\cal P}(x)$ describing the probabilistic shape of response $y$ at given push $x$. In Fig.~3 we plot three cross-sections of
the surface $\log {\cal P} (x,y)$ corresponding to $x= \$ \,\, 0.01, 0.07$ and $0.25$.
\begin{figure}[h]
 \begin{center}
 \epsfig{file=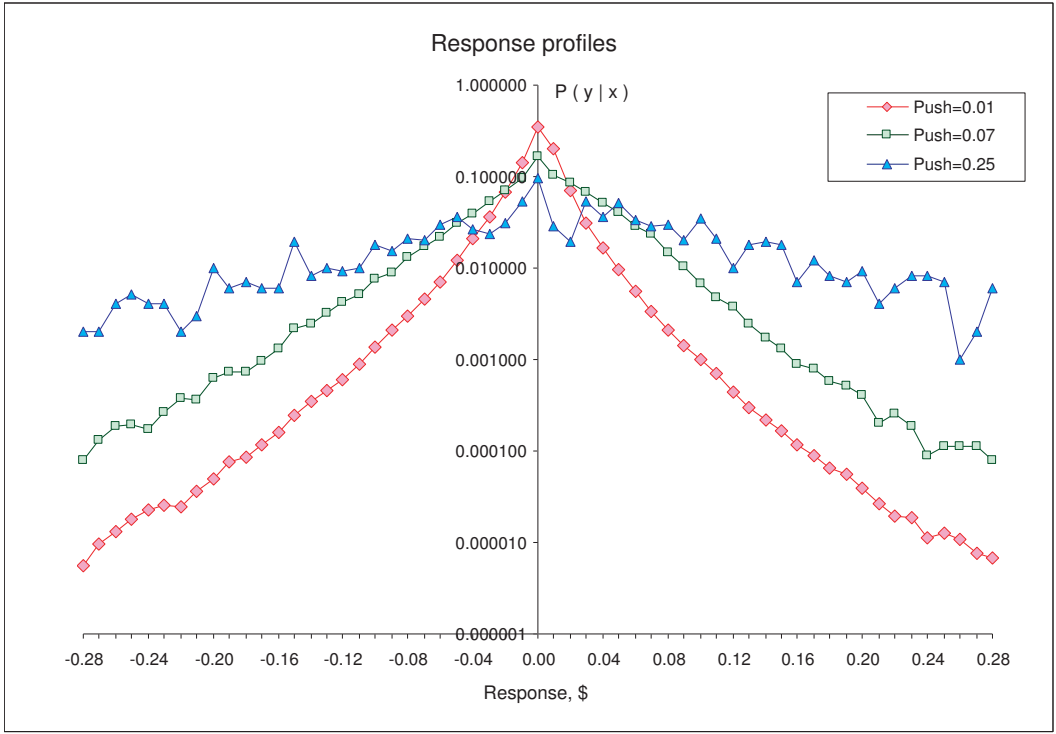,width=14cm}
 \end{center}
 \caption{Response profiles of ${\cal P}(x,y)$ for $x= \$ \, 0.01, \, 0.07$ and $0.25$}
\end{figure}
We observe a clear change in the structure of the response with growing push. Qualitatively this change can be described by evolution of the
parameter $\alpha$  in the stretched exponential distribution ${\cal P}_{\alpha (x)} (y) = {\cal N} (\alpha (x)) \exp \left[ - (|y|/\sigma)^{\alpha
(x)}/ \alpha (x) \right]$ from  $\alpha (x) \sim 1$ at small $|x|$ to $\alpha (x) \sim 2$ at large $|x|$, so that the distributions looks evolving
from the squeezed tent-like at small pushes to almost gaussian at large ones. Note that this interpretation is consistent with the suggested
description of the geometry of equiprobablity lines in Eq.~(\ref{equil}).

\subsubsection{Moments of conditional distribution}

Let us first consider a useful quantitative characteristics of a shape of the distribution ${\cal P}(y|\,x)$, the  conditional mean absolute response
\begin{equation}\label{mar}
 \langle | \, y| \rangle_x \, = \, \int dy | \, y| \, {\cal P} (y|\,x) \, .
\end{equation}
The dependence of  $\langle |y| \rangle_x$ on the push $x$ is plotted in Fig.~4.
\begin{figure}[h]
 \begin{center}
 \epsfig{file=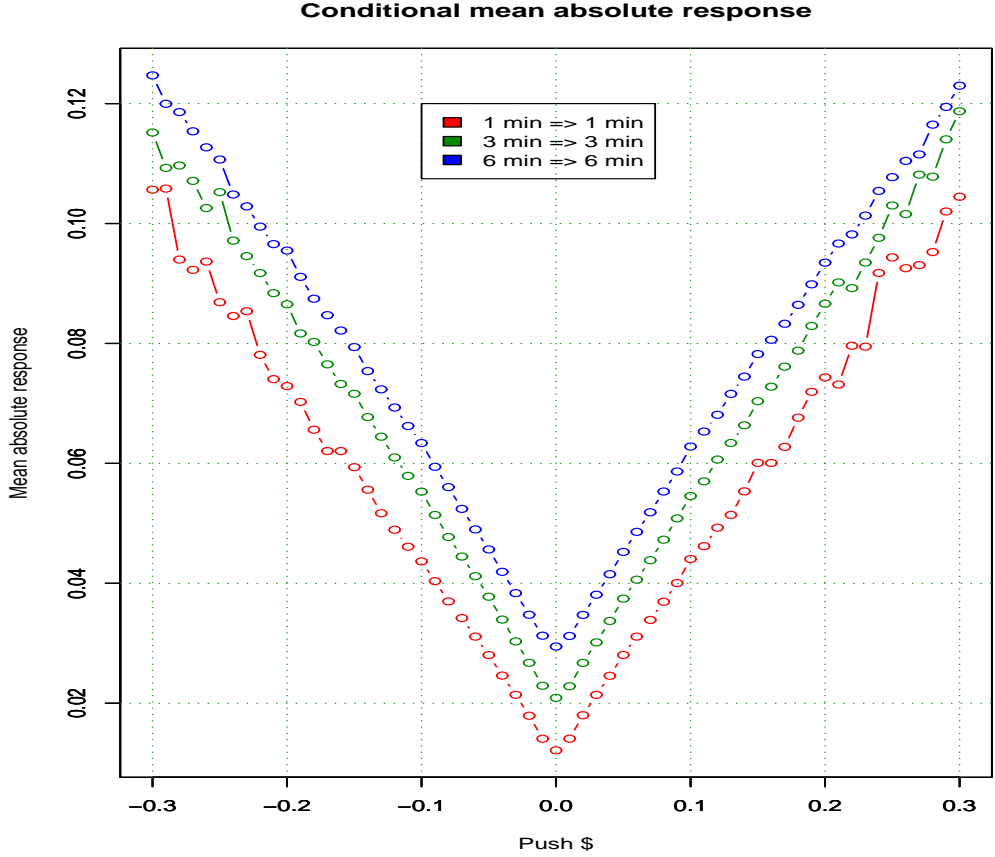,height=12cm,width=14cm}
 \end{center}
 \caption{The conditional mean of absolute increment versus the initial push}
\end{figure}
We see that to a good accuracy the mean absolute response is {\it linear} in the absolute value of the push, $\langle |y| \rangle_x \propto c_0 + c_1
\, |x|$. Let us recall that the mean response $\langle y \rangle_x$ is a {\it nonlinear} function of the push $x$ \cite{LTZZ05a}. As the absolute
response is a robust measure of volatility, in Fig.~4 we have an example of conditional volatility smile or dependence volatility smile that was
studied, in terms of a standard deviation of normalized returns, in \cite{LTZ05}. The dependence of  $\langle |y| \rangle_x$ on $x$ describes how
much of response volatility is created for a given push.

Because of the sensitivity of the mean conditional absolute response to the higher moments of the conditional distribution one can, by comparing it
to the value obtained for the Gaussian distribution with the same standard deviation, gauge the deviation of the distribution in question from the
Gaussian \cite{BP}. Let us thus consider the following ratio:
\begin{equation}\label{rhodef}
 \rho_x \, = \,  \frac{\langle |y| \rangle_x}{\langle |y| \rangle_x^G} \, ,
\end{equation}
where $\langle |y| \rangle_x^G = \sqrt{2/\pi} \sigma_x$ and $\sigma_x$ is an observed standard deviation of the corresponding conditional response.
The ratio $\rho_x$ is plotted, in three cases of consecutive 1 - minute, 3 - minute and 6 - minute intervals in Fig.~5
\begin{figure}[h]
 \begin{center}
 \epsfig{file=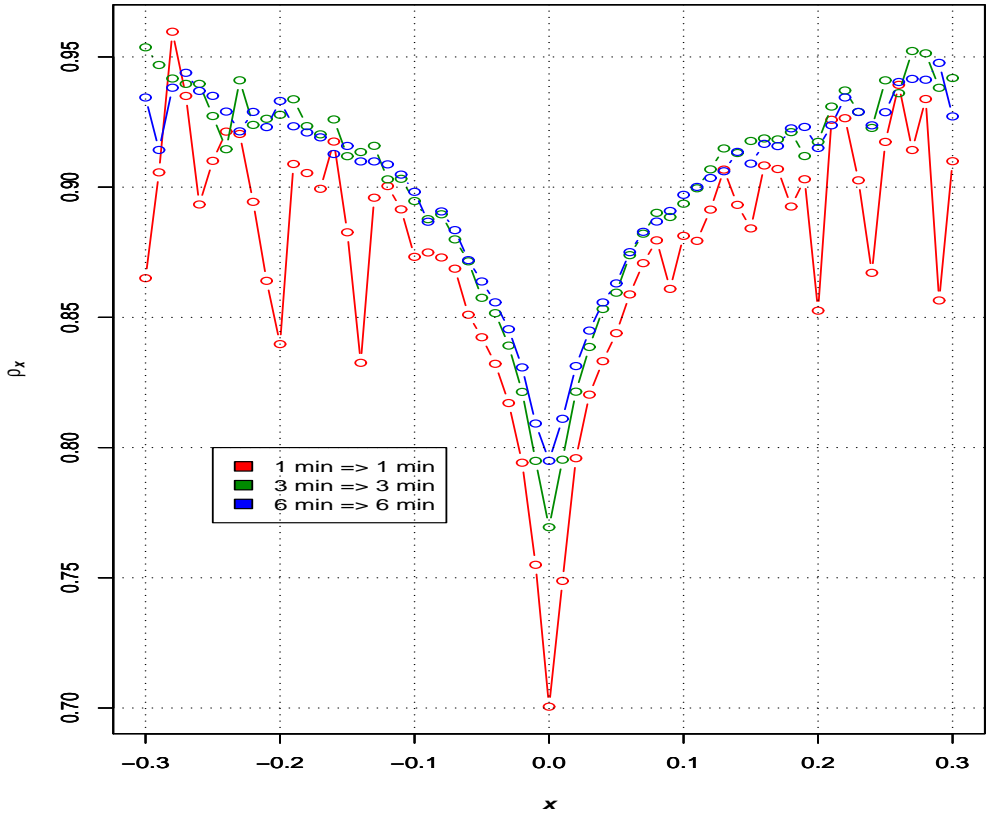,height=12cm,width=14cm}
 \end{center}
 \caption{Relative distance from the Gaussian distribution $\rho_x$}
\end{figure}
The pattern seen in Fig.~5 unambiguously shows the progressive "gaussization" of the response profile with growing push. This is a highly nontrivial
property of the distribution ${\cal P} (x,y)$. The same question can be addressed by computing the anomalous kurtosis of the conditional
distribution. The results obtained for anomalous kurtosis support the conclusion on "gaussization" but are rather noisy, especially for the case of 1
minute intervals.

The mean absolute response by default characterizes the symmetric component of the conditional probability distribution ${\cal P}(y|\,x)$ with
respect to the axis $y=0$. Asymmetry of ${\cal P}(y|\,x)$ with respect to this axis is characterized by its odd moments. The first moment, the mean
conditional response, was studied in detail in \cite{LTZZ05a}. It was shown that the mean conditional response has a nontrivial zizgag-shaped
dependence on the push. To probe the asymmetric contributions of higher order, let us consider the skewness ${\cal V}(x)$ of the conditional
response\footnote{Here we assume that the third moment of the conditional distribution exists.}
\begin{equation}
 {\cal V}(x) \, = \, \frac{\int dy \left( y-\langle y \rangle_x \right)^3 {\cal} P(y|\,x) }{ (\sigma^y_x)^3 \,} ,
\end{equation}
where $\sigma^y_x$ is the conditional standard deviation of response at given push. In Fig.~6 we plot the skewness of the response for the same set
of consecutive time intervals.
\begin{figure}[h]
 \begin{center}
 \epsfig{file=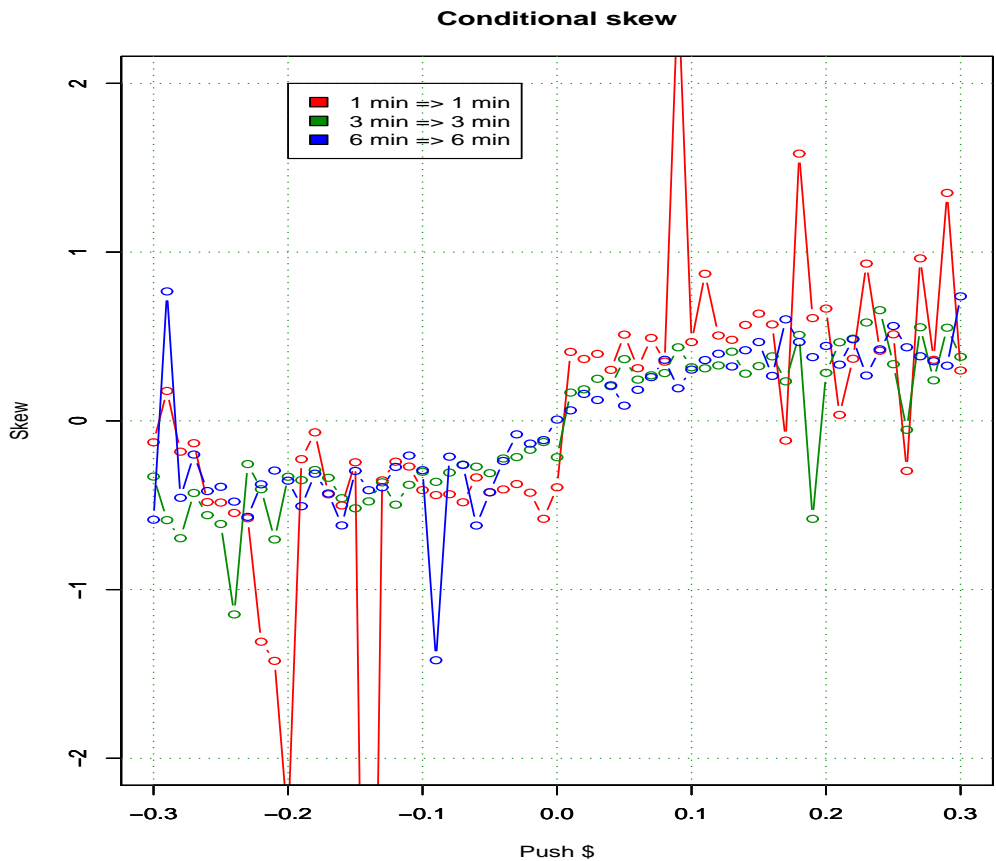,height=12cm,width=14cm}
 \end{center}
 \caption{The conditional skewness versus the initial push}
\end{figure}
The pattern seen in Fig.~6 corresponds to an interesting phenomenon. The asymmetry of the distribution of conditional response characterized by
skewness has the sign of the initial push, so that for negative pushes the response distribution is always negatively skewed, etc. Let us note that
the same conclusion can be reached by considering a robust characteristics of distribution asymmetry, a difference between the median of the
distribution and its mean.

The generic symmetry properties of the distribution ${\cal P}(x,y)$ are best revealed  by considering its symmetry with respect to the axes $y=0, \,
y=x, \, x=0$ and $y=-x$ (patterns I -- IV) \cite{LTZZ05a}. The patterns are of two types: pattern I is equivalent to pattern III and pattern II is
equivalent to pattern IV. Pattern I was analyzed above, so let us consider pattern II. To analyze the symmetry properties of the distribution ${\cal
P}(x,y)$ with respect to the axis $y=x$ it is convenient to introduce the new variables
\begin{equation}
 z = x+y \,\,\,\,\, {\rm and} \,\,\,\,\,\ {\tilde z}=y-x
\end{equation}
so that in this case one deals with the conditional distribution ${\cal P}({\tilde z}|z)$.

The dependence of the shape of ${\cal P}({\tilde z}|z)$ on the "push" $z$ can again be explored by considering the ratio
\begin{equation}
 \rho_z \, = \,  \frac{\langle |{\tilde z}| \rangle_z}{\langle |{\tilde z}| \rangle_z^G}
 \equiv \frac{\langle |\, y-x| \rangle_{x+y}}{\langle |\, y-x| \rangle_{x+y}^G}
\end{equation}
The ratio $\rho_z$ is plotted in Fig.~7.
\begin{figure}[h]
 \begin{center}
 \epsfig{file=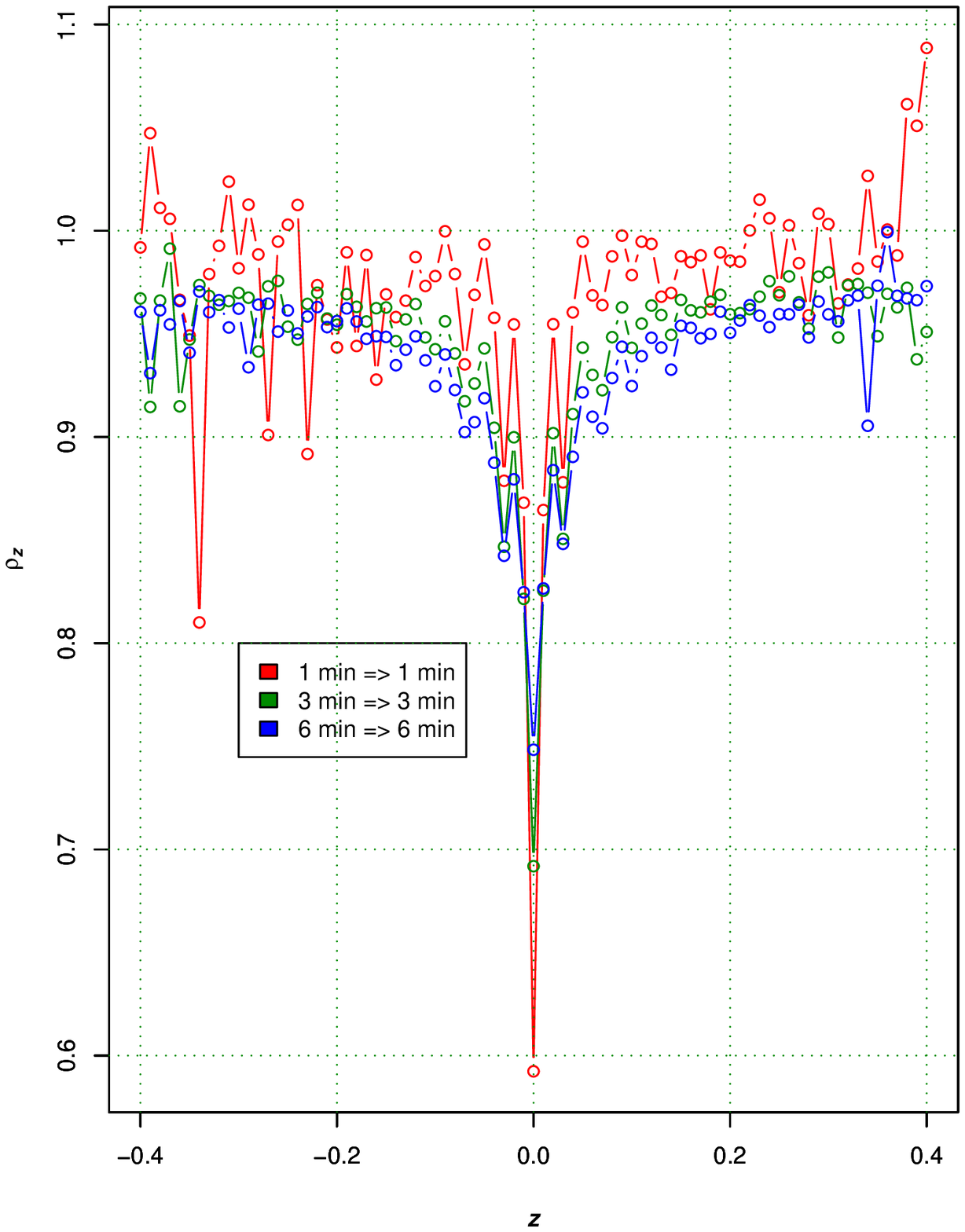,height=12cm,width=14cm}
 \end{center}
 \caption{Relative distance from the Gaussian distribution in rotated coordinates $\rho_z$}
\end{figure}
The generic pattern is the same as in Fig.~5: the distribution  ${\cal P}({\tilde z}|z)$ is progressively more and more gaussian with growing $|z|$.
The shape is somewhat different though, so that the gaussization process is different in this case.

\section{Discussion}

In the previous section we discussed a number of properties characterizing the probabilistic dependence patterns relating stock price increments in
consecutive time intervals. Our approach is based on the direct analysis of the bivariate probability distribution dependent on the two price
increments in question ${\cal P}(x,y)$ and of the corresponding conditional distribution ${\cal P}(y|\,x)$. We have concentrated on high frequency
data with increments in time intervals of length $\Delta T=$ 1 min., 3 min. and 6 min\footnote{The results of \cite{LTZZ05a} show that it is
reasonable to expect that all the features discussed in the present paper will hold at larger intraday time scales as well.}.

\subsection{Gaussization of conditional distribution at large magnitudes of price increments}

Let us first discuss in some more details a remarkable property of gaussization of multivariate distributions of price increments far away from the
origin in the $xy$ plane. A first hint comes from the analysis of the geometry of equiprobability levels of the push - response bivariate probability
distribution ${\cal P}(x,y)$. As seen in Fig.~1 and sketched in Fig.~2, the geometry of equibrobability lines changes from rhomboid in the vicinity
of the origin to circular far away from it, which is consistent with distribution changing the shape from bivariate Laplace to bivariate gaussian
one. This is, however, not a proof of gaussization: a Laplace distribution with standard deviation of ${\cal P}(y|x)$ growing with $| x|$ leads to
the same visual pattern.

Quantitative proof of gaussization comes from considering the ratio $\rho_x$ defined in eq.~(\ref{rhodef}) characterizing the shape of the
conditional response distribution. In this ratio the effect of variable conditional standard deviation is factored out. We have checked the
gaussization of the bivariate distribution ${\cal P}(x,y)$ along the two axes, $y=0$ and $y=x$ (see Figs.~5 and 7). The speed of gaussization turns
out different, but the phenomenon itself is undoubtedly present in both cases.

\subsection{Conditional dynamics: direct analysis of multivariate distribution vs regression models}

Knowledge of full bivariate distribution fully specifies corresponding conditional distributions and, therefore, a particular variant of conditional
dynamics. As already mentioned in the introduction, the main body of research on conditional dynamics in financial markets was done within a paradigm
of regression models \cite{SH,TL80,E82,EB86,B86,B87,H94,HS99,JR00,LL02,CD04}, in which conditional distribution for the forthcoming price increment
depends on lagged increments and lagged conditional moments. In the simplest version of the AR(1)-ARCH(1) model \cite{SH,E82} the conditional
distribution of the return $r_y$ is gaussian, ${\rm Law}(r_y)={\cal P}_G (r_y |\, \mu_y, \sigma_y)$, where conditional mean $\mu_y$ and conditional
standard deviation $\sigma_y$ depend on return $r_x$ in the previous interval, $\mu_x \propto r_x$ and $\sigma^2_y = \alpha + \beta \, r_x^2$. This
setting is methodologically equivalent to the one considered in the present paper in the sense that the only information required for computing the
conditional distribution for $r_y$ is the value of $r_x$, so that the conditional dynamics of AR(1)-ARCH(1) is fully described by conditional
distribution ${\cal P}(r_y|\,r_x)$. To match the features described in the preceding \cite{LTZZ05a} and present papers one would have to consider a
nonlinear dependence of $\mu_y$ on $r_x$ and construct a fairly complicated fat-tailed skewed conditional probability distribution. Even forgetting
for a moment about the z-shaped dependence of mean conditional response on the push, meaningful comparison could be with a version of ARCH(1) with a
fat-tailed skewed conditional distribution. As to the zigzag-shaped dependence of the mean conditional response on the push, the most natural
treatment could be given within a class of threshold autoregressive (TAR) models \cite{TL80}.

To compare our results with the properties of autoregressive models considered in the literature let us consider the AR(1)-GARCH(1,1) model
\cite{SH,EB86}, particularly its versions with fat-tailed \cite{B87} and conditionally fat-tailed and skewed conditional distributions for residuals
\cite{H94,HS99,JR00,LL02,CD04}. In these models conditional volatility is a function of both lagged returns and volatility, so a comparison with the
results obtained using the bivariate distribution ${\cal P}(x,y)$ is not direct. With this remark in mind, let us make some comparison at the "moment
by moment" basis.
\begin{itemize}
 \item{
        The conditional mean in the AR(1) model is by default a linear function of the push, $\mu_x \propto r_x$. This is equivalent to
        an assumption of the ellipticity of the underlying bivariate distribution ${\cal P}(r_x,r_y)$. As shown in \cite{LTZZ05a}, the intraday
        conditional dynamics is characterized by a fairy complex nonlinear z-shaped pattern of mean conditional response. To take this into
        account one should generalize AR(1) to TAR(1).
      }
 \item{
        The conditional standard deviation $\sigma_y$ in GARCH(1,1) models is a growing function of $r_x$,
        $\sigma^2_y = {\rm const.}+ r^2_x + \varepsilon_\sigma$. This is consistent with the dependence smile discussed in \cite{LTZ05}
        and shown in Fig. 3.
      }
 \item{
        The results obtained in the framework of generalized GARCH models for the conditional skewness \cite{H94,HS99,JR00,LL02,CD04} (only the
        daily timescale was considered) are somewhat contradictory. In \cite{JR01,HS99,JR00,LL02} a conclusion was that negative return is followed
        by negative conditional skew - in agreement with the results of the previous section, while no conclusion on the sign of conditional skew
        following the positive return was reached. At the same time, the conclusion of \cite{CD04} was that the sign of conditional skew is always
        opposite to the sign of initial return.
      }
 \item{
        The main result on conditional kurtosis obtained within the generalized GARCH approach was \cite{JR00} that it is time dependent and not
        always existent. A comparison with our result on the progressive gaussization of the response distribution with growing magnitude of the
        push does not seem possible.
      }
\end{itemize}

Our method of directly analyzing the conditional distribution allows to describe its properties in a model-independent framework. If analyzed from
the point of view of regressive conditional dynamics the results described in the present paper and in the preceding paper \cite{LTZZ05a} can be
formulated as follows. The conditional dynamics is
\begin{itemize}
\item{ nonlinear}
\item{ heteroskedastic as seen in the volatility dependence smile}
\item{ not conditionally gaussian}
\item{ characterized by conditional skew depending on the lagged increment}
\item{ characterized by conditional fat-tailedness that diminishes with growing magnitude of lagged increment}
\end{itemize}

\section{Conclusions and outlook}

Let us formulate once again the main conclusions of the paper. Studying the geometry of the full bivariate probability distribution ${\cal P}(x,y)$
and the corresponding conditional distribution ${\cal P}(y|\,x)$ we have found that
 \begin{itemize}
 \item{The shape of equiprobability lines of the bivariate probability distribution ${\cal P}(x,y)$ changes from roughly rhomboid
 in the vicinity of the origin to roughly circular far away from it.}
 \item{The conditional distribution ${\cal P}(y|\,x)$ is shown to become progressively more gaussian at increasing push magnitudes. Analogous
  gaussization takes place for conditional distribution considered with respect to the axis $y=x$.}
 \item{The bivariate distribution ${\cal P}(x,y)$ is approximately invariant with respect to rotations at multiples of $\pi/2$}
 \item{The conditional mean absolute response  is linear in the absolute value of push}
 \item{The skewness of the response distribution inherits a sign of the push}
 \end{itemize}

As was emphasized above in \cite{LTZZ05a} and the present paper we study a combined ensemble of all pairs of consecutive price increments from all
stocks. How is this overall geometry related to the geometric properties of individual stocks? Only after answering this question can one come close
to describing the microscopic mechanism underlying the uncovered probabilistic dependence patterns. This issue is analyzed in the companion paper
\cite{LTZZ05b}.

The work of A.L. was partially supported by the Scientific school support grant 1936.2003.02

\section{Appendix}

Below we list stocks studied in the paper:

\medskip

 A, AA, ABS, ABT, ADI, ADM, AIG, ALTR, AMGN, AMD, AOC, APA, APOL, AV, AVP, AXP,
 BA, BBBY, BBY, BHI, BIIB, BJS, BK, BLS, BR, BSX,
 CA, CAH, CAT, CC, CCL, CCU, CIT, CL, COP, CTXS, CVS, CZN,
 DG, DE,
 EDS, EK, EOP, EXC,
 FCX, FD, FDX, FE, FISV, FITB, FRE,
 GENZ, GIS,
 HDI, HIG, HMA, HOT, HUM,
 JBL, JWN,
 INTU,
 KG, KMB, KMG,
 LH, LPX, LXK,
 MAT, MAS, MEL, MHS, MMM, MO, MVT, MX, MYG,
 NI, NKE, NTRS,
 PBG, PCAR, PFG, PGN, PNC, PX,
 RHI, ROK,
 SOV, SPG, STI, SUN,
 T, TE, TMO, TRB, TSG,
 UNP, UST,
 WHR, WY


\begin{thebibliography}{99}

\bibitem{LTZZ05a}
A.~Leonidov, V.~Trainin, S.~Zaitsev, A.~Zaitsev, "Market Mill Dependence Pattern in the Stock Market: Asymmetry Structure, Nonlinear Correlations and
Predictability", arXiv:physics/0601098.

\bibitem{BDS87}
W.~Brock, W.~Dechert, J.~Scheinkman, "A test of independence based on the correlation dimension", Working paper, University of Wisconsin at Madison,
University of Houston and University of Chicago (1987).

\bibitem{H91}
D.H.~Hsieh, "Chaos and Nonlinear Dynamics: Application to Financial Marekts", {\it Journal of Finance} {\bf 46} (1991), 1839-1878

\bibitem{Lo}
A.C.~MacKinlay, A.W.~Lo, J.Y.~Kampbell, {\it The Econometrics of Financial Markets}, Princeton, 1997; \\ A.W.~Lo, A.C.~MacKinlay, {\it A Non-Random
Walk Down Wall Sreet}, Princeton, 1999

\bibitem{Man}
B.~Mandelbrot, "Fractal and Multifractal Finance. Crashes and Long-dependence", www.math.yale.edu/mandelbrot/webbooks/wb\_fin.html

\bibitem{SH}
A.~Shiryaev, "Essentials of Stochastic Finance: Facts, Models, Theory", World Scientific, 2003

\bibitem{BP}
J.-P.~Bouchaud, M.~Potters, {\it Theory of Financial Risk and Derivative Pricing}, Cambridge, 2000, 2003.



\bibitem{LTZ05}
A.~Leonidov, V.~Trainin, A.~Zaitsev, "On collective non-gaussian dependence patterns in high frequency financial data", ArXiv:physics/0506072,
submitted to {\it Quantitative Finance}

\bibitem{LTZZ05b}
A.~Leonidov, V.~Trainin, A.~Zaitsev, S.~Zaitsev, "Market Mill Dependence Pattern in the Stock Market: Individual Portraits", in preparation


\bibitem{M63}
B.~Mandelbrot, "The Variation of Certain Speculative Prices", {\it Journal\ of\ Business}\ {\bf 36} (1963), 394-419

\bibitem{MB}
B.~Mandelbrot, R.L.~Hudson, "The (Mis)behavior of Prices:  A Fractal View of Risk, Ruin, and Reward". New York: Basic Books; London: Profile Books,
2004

\bibitem{CL96}
T.F.~Crack, O.~Ledoit, "Robust Structure Without Predictability: The "Compass Rose" Pattern of the Stock Market", {\it The Journal of Finance}\ {\bf
51} (1996), 751-762

\bibitem{AV04}
A.~Antoniou, C.E.~Vorlow, "Price Clustering and Discreteness: Is there Chaos behind the Noise?", arXiv:cond-mat/0407471

\bibitem{V04}
C.E.~Vorlow, "Stock Price Clustering and Discreteness: The "Compass Rose" and Predictability", arXiv:cond-mat/0408013

\bibitem{EMS99}
P.~Embrechts, A.~McNeil, D.~Straumann, "Correlation and depenendence in risk management: properties and pitfalls", Risk Lab working paper (1999)

\bibitem{ELM01}
P.~Embrechts, P.~Lindskog, A.~McNeil, "Modelling Dependence with Copulas and Applications to Risk Management", Risk Lab working paper (2001)

\bibitem{MS01}
Y.~Malevergne, D.~Sornette, "Testing the Gaussian Copula Hypothesis for Financial Asset Dependencies", ArXix:cond-mat/0111310

\bibitem{JR01}
E.~Jondeau, M.~Rockinger, "Conditional Dependency of Financial Series: an Application of Copulas", Banque de France working paper NER 82 (2001)

\bibitem{CJY05}
K.~Chen, C.~Jayprakash, B.~Yuan, "Conditional Probability as a Measure of Volatility Clustering in Financial Time Series", arXiv:physics/0503157

\bibitem{BM03}
M.~Boguna, J.~Masoliver, "Conditional dynamics driving financial markets", ArXiv:cond-mat/0310217


\bibitem{TL80}
H.~Tong, K.~Lim, "Threshold autoregression, limit cycles and cyclical data", {\it Journal of the Royal Statistical Society}\, {\bf B42} (1980),
245-292

\bibitem{E82}
R.F.~Engle, "Autoregressive Conditional Heteroskedasticity with Estimates of the Variance of United Kingdom Inflation", {\it Econometrica}\ {\bf 50}
(1982), 987-1007

\bibitem{EB86}
R.F.~Engle, T.~Bollerslev, "Nodelling the persistence of conditional variances", {\it Economics Reviews}\ {\bf 5} (1986), 1-50

\bibitem{B86}
T.~Bollerslev, "Generalized autoregressive conditional heteroskedasticity", {\it Journal of Econometrics}\ {\bf 31} (1986), 307-327

\bibitem{B87}
T.~Bollerslev, "A Conditionally Heteroscedastic Time Series Model For Speculative Prices Rates Of Return", {\it Review of Economics and Statistics}\
{\bf 69} (1987), 542-547

\bibitem{H94}
B.E.~Hansen, "Autoregressive Conditional Density Estimation", {\it International Economic Review} {\bf 35} (1994), 705-730

\bibitem{HS99}
C.R.~Harvey, A.~Siddique, "Autoregressive Conditional Skewness", {\it Journal of Finanical and Quantitative Analysis}\ {\bf 34} (1999), 465-487

\bibitem{JR00}
E.~Jondeau, M.~Rockinger, "Conditional volatility, skewness, and kurtosis: existence and persistence", Banque de France working paper NER 77 (2000)

\bibitem{LL02}
P.~Lambert, S.~Laval, "Modelling skewness dynamics in series of financial data using skewed location-scale distributions", University of Louvain
working paper (2002)

\bibitem{CD04}
A.~Charoenrook, H.~Daouk, "Conditional Skewness of Aggregate Market Returns", NBER working paper (2004)







\end{thebibliography}
\end{document}